\documentclass[aps,preprint,tightenlines,a4paper,showpacs,nofootinbib]{revtex4}
\usepackage{epsfig}
\usepackage{multirow}
\usepackage{amsmath,stackrel}
\usepackage{graphicx}
\begin{document}
\title{Heavy flavor dibaryons}
\author{T.~F.~Caram\'es}
\affiliation{Departamento de F{\'\i}sica Fundamental and IUFFyM,
Universidad de Salamanca, 37008 Salamanca, Spain}
\author{A.~Valcarce}
\affiliation{Departamento de F{\'\i}sica Fundamental and IUFFyM,
Universidad de Salamanca, 37008 Salamanca, Spain}
\date{\emph{Version of }\today}

\begin{abstract}
We study the two-baryon system with two units of charm looking for the 
possible existence of a loosely bound state resembling the 
$H$ dibaryon. We make use of a chiral constituent quark model 
tuned in the description of the baryon and meson spectra 
as well as the $NN$ interaction. The presence of the heavy 
quarks makes the interaction simpler than in light baryon systems.
We analyze possible quark-Pauli effects that would be present
in spin-isospin saturated channels. Our results point to the 
non-existence of a charmed $H$-like dibaryon, although it may
appear as a resonance above the $\Lambda_c\Lambda_c$ threshold.
\end{abstract}

\maketitle

\section{Introduction}
\label{secI}

Hadronic molecules have recently become a hot topic on the study of
meson spectroscopy as well-suited candidates for the structure of the so-called
XYZ states~\cite{Esp14}. This explanation seems compulsory in the case of the
charged charmonium-like states first reported by Belle~\cite{Cho08}
and charged bottomonium-like states reported later on also by Belle~\cite{Bon12}. 
In the two-baryon system there is a well-established molecule composed of two light baryons,
the deuteron. Another well-known candidate is the $H$ dibaryon suggested 
by Jaffe~\cite{Jaf77}, a bound state with quantum numbers of the 
$\Lambda\Lambda$ system, i.e., strangeness $-2$ and $(T)J^P=(0)0^+$.
 
These two scenarios have recently triggered several studies about the
possible existence of molecules made of heavy 
baryons~\cite{Lee11,Meg11,Liz12,Oka13,Hua14}. The main
motivation originates from the reduction of the kinetic energy due
to large reduced mass as compared to systems made of light baryons.
However, such molecular states that have been intriguing objects of 
investigations and speculations for many years, are usually the concatenation of
several effects and not just a fairly attractive interaction. The coupling 
between close channels or the contribution of non-central forces used to 
play a key role for their existence. Some of these contributions may be 
reinforced by the presence of heavy quarks while others will become weaker. 

The understanding of the hadron-hadron interaction is an important 
topic nowadays. To encourage new experiments seeking for evidence 
of theoretical predictions, it is essential to make a 
detailed theoretical investigation of the possible existence of 
bound states, despite some uncertainty in contemporary interaction 
models~\cite{Ric15}. It is the purpose of this work to analyze the
interaction of two-baryons with two units of charm and its application
to study the possible existence of a hadronic molecule with the
quantum numbers of the $\Lambda_c\Lambda_c$ system, $(T)J^P=(0)0^+$.
When tackling this problem, one has to manage with an important difficulty, namely the 
complete lack of experimental data. Thus, the generalization 
of models describing the two-hadron interaction in the light flavor sector could offer 
insight about the unknown interaction of hadrons with heavy flavors. 
Following these ideas, we will make use of a chiral constituent quark
model (CCQM) tuned on the description of the $NN$ interaction~\cite{Val05} as 
well as the meson~\cite{Vij05} and baryon~\cite{Vag05,Val08} spectrum in all flavor sectors,
to obtain parameter-free predictions that may be testable in 
future experiments. Such a project was already undertaken for the 
interaction between charmed mesons with reasonable predictions~\cite{Car09},
what encourages us in the present challenge. Let us note that the study of 
the interaction between charmed baryons has become an interesting subject 
in several contexts~\cite{Pan09} and it may shed light on the possible 
existence of exotic nuclei with heavy flavors~\cite{Dov77}. 

The paper is organized as follows. We will use Sec.~\ref{secII} for describing
all technical details of our calculation. In particular, Sec.~\ref{secIIA} 
presents the description of the two-baryon system quark-model wave function, 
analyzing its short-range behavior looking for quark-Pauli effects. Sec.~\ref{secIIB}
briefly reviews the interacting potential, and section~\ref{secIIC} deals with
the solution of the two-body problem by means of the Fredholm determinant.
In Sec.~\ref{secIII} we present our results. We will discuss the baryon-baryon 
interactions emphasizing those aspects that may produce different results from
purely hadronic theories. We will analyze the character of the interaction in the 
different isospin-spin channels, looking for the attractive ones that may lodge 
resonances. We will also compare with existing results in the literature.
Finally, in Sec.~\ref{secIV} we summarize our main conclusions.

\section{The two-baryon system}
\label{secII}
\subsection{The two-baryon wave function}
\label{secIIA}
\begin{figure}[b]
\vspace*{-4cm}
\includegraphics[width=14cm]{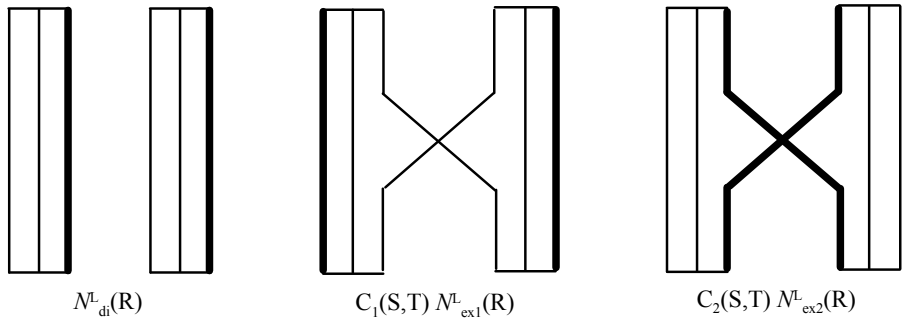}
\vspace*{-10cm}
\caption{Different diagrams contributing to the two-baryon normalization
kernel as indicated in Eq.~(\ref{Norm}). The vertical thin solid lines represent a light quark, $u$ or $d$,
while the vertical thick solid lines represent the charm quarks.}
\label{fig1}
\end{figure}
We describe the baryon-baryon system by means of a constituent quark
cluster model, i.e., baryons are described as clusters of three constituent
quarks. Assuming a two-center shell model the wave function of
an arbitrary baryon-baryon system can be written as:
\begin{eqnarray}
\Psi_{B_1 B_2}^{{\rm LST}}({\vec R}) & = & {\frac{\cal A}{\sqrt{1 +
\delta_{B_1 B_2}}}} \sqrt{\frac{1}{2}} \Biggr\{ \left[
\Phi_{B_1} \left( 123 ;{-\frac{\vec R}{2}} \right)
\Phi_{B_2} \left( 456 ; {\frac{\vec R}{2}} \right)
\right]_{{\rm LST}} \, + \nonumber \\
& + &(-1)^{f} \,
\left[
\Phi_{B_2} \left( 123 ; {-\frac{\vec R}{2}} \right)
\Phi_{B_1} \left( 456 ; {\frac{\vec R}{2}} \right)
\, \right]_{{\rm LST}} \Biggr\} \,
,\label{Gor}
\end{eqnarray}
where ${\cal A}$ is the antisymmetrization operator accounting for 
the possible existence of identical quarks inside the hadrons. The symmetry 
factor $f$ satisfies $L+S_1+S_2-S+T_1+T_2-T+f=$ odd~\cite{Val05}. For non-identical baryons 
indicates the symmetry associated to a given set of values LST. The non-possible 
symmetries correspond to forbidden states. For identical baryons, $B_1=B_2$,
$f$ has to be even in order to have a non-vanishing wave function, recovering 
the well-known selection rule
$L+S+T=$ odd. In the case we are
interested in, two baryons with a charmed quark, the antisymmetrization operator comes given by
\begin{equation}
{\cal A} = \left ( 1-\sum_{i=1}^2 \sum_{j=4}^5 P^{LST}_{ij} - P^{LST}_{36} \right )(1-{\cal P}) \label{ant} \, ,
\end{equation}
where $P^{LST}_{ij}$ exchanges a pair of identical quarks $i$ and $j$ and $\cal P$
exchanges identical baryons.
There appear two different contributions coming either from the exchange of light
quarks ($i=1,2$ and $j=4,5$) or the exchange of the charm quarks ($i=3$ and $j=6$).
If we assume gaussian $0s$ wave functions for the quarks inside the hadrons,
the normalization of the two--baryon wave function $\Psi_{B_i B_j}^{LST}({\vec R})$
of Eq.~(\ref{Gor}) can be expressed as,
\begin{equation}
{\cal N}_{B_iB_j}^{LST}(R)= {\cal N}^{L}_{\rm di}(R) - C_1(S,T) \, {\cal N}^{L}_{\rm ex1}(R)  - 
C_2(S,T) \, {\cal N}^{L}_{\rm ex2}(R)\, .
\label{Norm}
\end{equation}
${\cal N}_{\rm di}^{L}(R)$, ${\cal N}_{\rm ex1}^{L}(R)$, and ${\cal N}_{\rm ex2}^{L}(R)$ stand for
the direct and exchange radial normalizations depicted in Fig.~\ref{fig1},
whose explicit expressions are
\begin{eqnarray}
 {\cal{N}}_{\rm di}^{L} (R) &=& 4 \pi \exp \left\lbrace {-\frac{R^2}{8} 
\left( \frac{4}{b^2} + \frac{2}{b_c^2}\right)}\right\rbrace  i_{L+1/2} 
\left[ \frac{R^2}{8} \left( \frac{4}{b^2} + \frac{2}{b_c^2} \right)\right] \,, \\ \nonumber
{\cal{N}}_{\rm ex1}^{L} (R) &=& 4 \pi \exp \left\lbrace {-\frac{R^2}{8} 
\left( \frac{4}{b^2} + \frac{2}{b_c^2}\right)}\right\rbrace  i_{L+1/2} 
\left[ \frac{R^2}{8} \left( \frac{2}{b_c^2} \right) \right] \label{nex12} \,, \\
{\cal{N}}_{\rm ex2}^{L} (R) &=& 4 \pi \exp \left\lbrace {-\frac{R^2}{8} 
\left( \frac{4}{b^2} + \frac{2}{b_c^2}\right)}\right\rbrace  i_{L+1/2} 
\left[ \frac{R^2}{8} \left( \frac{4}{b^2} - \frac{2}{b_c^2} \right) \right] \,, \nonumber
\label{Norm2}
\end{eqnarray}
where, for the sake of generality, we have assumed different gaussian parameters
for the wave function of the light quarks ($b$) and the heavy quark ($b_c$).
In the limit where the two hadrons overlap ($R \to 0$), the Pauli principle
may impose antisymmetry requirements not present in a hadronic description.
Such effects, if any, will be prominent for $L=0$. Using the asymptotic form
of the Bessel functions, $i_{L+1/2}$, we obtain,
\begin{eqnarray}
{\cal{N}}_{\rm di}^{L=0} &\stackrel[R\to 0]{}{\hbox to 20pt{\rightarrowfill}}& 4 \pi \left\lbrace {1 -\frac{R^2}{8} 
\left( \frac{4}{b^2} + \frac{2}{b_c^2}\right)}\right\rbrace  
\left[1 + \frac{1}{6} \left( \frac{R^2}{8} \left( \frac{4}{b^2} + \frac{2}{b_c^2} \right) \right)^2 + ...\right] \nonumber\,, \\
{\cal{N}}_{\rm ex1}^{L=0} &\stackrel[R\to 0]{}{\hbox to 20pt{\rightarrowfill}}& 4 \pi \left\lbrace {1 -\frac{R^2}{8} 
\left( \frac{4}{b^2} + \frac{2}{b_c^2}\right)}\right\rbrace  
\left[1 + \frac{1}{6} \left( \frac{R^2}{8} \left( \frac{2}{b_c^2} \right)\right)^2 + ...\right] \,, \\ 
{\cal{N}}_{\rm ex2}^{L=0} &\stackrel[R\to 0]{}{\hbox to 20pt{\rightarrowfill}}& 4 \pi \left\lbrace {1 -\frac{R^2}{8} 
\left( \frac{4}{b^2} + \frac{2}{b_c^2}\right)}\right\rbrace  
\left[1 + \frac{1}{6} \left( \frac{R^2}{8} \left( \frac{4}{b^2} - \frac{2}{b_c^2} \right)\right)^2 + ...\right] \nonumber \,. \label{npro}
\end{eqnarray}
Finally, the normalization kernel of Eq.~(\ref{Norm}) can be written for the S wave in the overlapping region
as,
\begin{equation}
{\cal N}_{B_iB_j}^{L=0ST} \stackrel[R\to 0]{}{\hbox to 20pt{\rightarrowfill}} 4\pi \left\lbrace {1 -\frac{R^2}{8} 
\left( \frac{4}{b^2} + \frac{2}{b_c^2}\right)}\right\rbrace  
\left\lbrace \left[ 1 - 3 C(S,T) \right] + ... \right\rbrace \,,
\end{equation}
where $C(S,T)=\frac{4}{9} C_1(S,T) + \frac{1}{9} C_2(S,T)$ for systems with two charmed baryons.
For the particular case of the $N\Xi_{cc}$ system $C(S,T)=\frac{1}{3} C_1(S,T)$\footnote{Note that in 
the $N\Xi_{cc}$ case the antisymmetrization operator of Eq.~(\ref{ant}) becomes much more simpler
${\cal A} = \left ( 1- 3 P^{LST}_{ij}\right)$, and only the first exchange diagram of Fig.~\ref{fig1}
will contribute with the two charm quarks on the same baryon.}.
The values of $C(S,T)$ are given in Table~\ref{tab1}.
\begin{table}[b]
\caption{$C(S,T)$ spin-isospin coefficients, as defined in the text, for $L=0$ partial waves.}
\label{tab1}
\begin{tabular}{|c|ccc|ccc|c|}
\hline \hline
          &\multicolumn{3}{c|}{$T=0$}    &  \multicolumn{3}{c|}{$T=1$}     & $T=2$          \\
$B_i B_j$ & $\Lambda_c\Lambda_c$ & $N \Xi_{cc}$ & $\Sigma_c\Sigma_c$  & $N \Xi_{cc}$  & $\Lambda_c\Sigma_c$ & $\Sigma_c\Sigma_c$ & $\Sigma_c\Sigma_c$  \\ \hline
$J=0$     & $2/9$                & $-1/9$       & $0$                 & $5/27$        & $-1/18$             & $-$                & $-1/9$                \\
$J=1$     & $-$                  & $-2/27$      & $-$                 & $16/81$       & $1/54$              & $-5/81$            & $-$                \\
\hline\hline
\end{tabular}
\end{table}
Thus, the closer the value of $C(S,T)$ to 1/3 the larger the suppression of the normalization
of the wave function at short distances, generating Pauli repulsion. It is the channel 
$\Lambda_c \Lambda_c$ $(T,J)=(0,0)$, with $C(S,T)=2/9$, where
the norm kernel gets smaller at short distances. One would only find Pauli blocking~\cite{Val97}
in excited states like $\Sigma_c^* \Sigma_c ^*$ $(T,J)=(2,3)$, where $C(S,T)=1/3$,
due to lacking degrees of freedom to accommodate the light quarks present on this configuration,
four $u$ quarks with spin up. However, this partial wave
could only exist for $L=$ odd, and then the Pauli blocking may get
masked by the centrifugal barrier.

\begin{figure}[t]
\includegraphics[width=7cm]{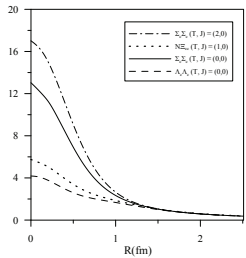}
\caption{Normalization kernel as defined in Eq.~(\protect{\ref{Norm}}) for $L=0$ and four different channels.}
\label{fig2}
\end{figure}
We show in Fig.~\ref{fig2} the normalization kernel given 
by Eq.~(\ref{Norm}) for $L=0$ and four different channels:
$\Lambda_c\Lambda_c$ with $(T,J)=(0,0)$, $\Sigma_c\Sigma_c$
with $(T,J)=(0,0)$ and $(2,0)$, and $N\Xi_{cc}$ with $(T,J)=(1,0)$. In the first case $C(S,T)$
is positive and close to $1/3$, what gives a small normalization kernel.
In the last case $C(S,T)$ is also positive but smaller, giving rise to a slightly larger 
normalization kernel. In the other two cases $C(S,T)$ is zero or negative, 
showing a large norm kernel at short distances 
and therefore one does not expect any Pauli effect at all.

\subsection{The two-body interactions}
\label{secIIB}

The interactions involved in the study of the two-baryon
system are obtained from a chiral constituent quark model~\cite{Val05}. 
This model was proposed in the early 90's in an attempt to
obtain a simultaneous description of the nucleon-nucleon
interaction and the light baryon spectra. It was later on generalized to all 
flavor sectors~\cite{Vij05}. 
In this model hadrons are described as clusters of three interacting 
massive (constituent) quarks, the mass coming from the spontaneous breaking 
of the original $SU(2)_{L}\otimes SU(2)_{R}$ 
chiral symmetry of the QCD Lagrangian.
QCD perturbative effects are taken into account
through the one-gluon-exchange (OGE) potential~\cite{Ruj75}.
It reads,  
\begin{equation}
V_{\rm OGE}({\vec{r}}_{ij})=
        {\frac{\alpha_s}{4}}\,{\vec{\lambda}}_{i}^{\rm
c} \cdot {\vec{\lambda}}_{j}^{\rm c}
        \Biggl \lbrace{ \frac{1}{r_{ij}}}
        - \dfrac{1} {4} \left(
{\frac{1}{{2\,m_{i}^{2}}}}\, +
{\frac{1}{{2\,m_{j}^{2}}}}\,
        + {\frac{2 \vec \sigma_i \cdot \vec
\sigma_j}{3 m_i m_j}} \right)\,\,
          {\frac{{e^{-r_{ij}/r_{0}}}}
{{r_{0}^{2}\,\,r_{ij}}}}
        - \dfrac{3 S_{ij}}{4 m_i m_j r_{ij}^3}
        \Biggr \rbrace\,\, ,
\end{equation}
where $\lambda^{c}$ are the $SU(3)$ color matrices, 
$r_0=\hat r_0/\mu$ is a flavor-dependent regularization scaling with the 
reduced mass of the interacting pair, and $\alpha_s$ is the
scale-dependent strong coupling constant given by~\cite{Vij05},
\begin{equation}
\alpha_s(\mu)={\frac{\alpha_0}{\rm{ln}\left[{({\mu^2+\mu^2_0})/
\gamma_0^2}\right]}},
\label{asf}
\end{equation}
where $\alpha_0=2.118$, 
$\mu_0=36.976$ MeV and $\gamma_0=0.113$ fm$^{-1}$. This equation 
gives rise to $\alpha_s\sim0.54$ for the light-quark sector,
$\alpha_s\sim0.43$ for $uc$ pairs, and
$\alpha_s\sim0.29$ for $cc$ pairs.

Non-perturbative effects are due to the spontaneous breaking of the original 
chiral symmetry at some momentum scale. In this domain of momenta, light quarks 
interact through Goldstone boson exchange potentials,
\begin{equation}
V_{\chi}(\vec{r}_{ij})\, = \, V_{\rm OSE}(\vec{r}_{ij}) \, + \, V_{\rm OPE}(\vec{r}_{ij}) \, ,
\end{equation}
where
\begin{eqnarray}
V_{\rm OSE}(\vec{r}_{ij}) &=&
    -\dfrac{g^2_{\rm ch}}{{4 \pi}} \,
     \dfrac{\Lambda^2}{\Lambda^2 - m_{\sigma}^2}
     \, m_{\sigma} \, \left[ Y (m_{\sigma} \,
r_{ij})-
     \dfrac{\Lambda}{{m_{\sigma}}} \,
     Y (\Lambda \, r_{ij}) \right] \,, \nonumber \\
V_{\rm OPE}(\vec{r}_{ij})&=&
     \dfrac{ g_{\rm ch}^2}{4
\pi}\dfrac{m_{\pi}^2}{12 m_i m_j}
     \dfrac{\Lambda^2}{\Lambda^2 - m_{\pi}^2}
m_{\pi}
     \Biggr\{\left[ Y(m_{\pi} \,r_{ij})
     -\dfrac{\Lambda^3}{m_{\pi}^3} Y(\Lambda
\,r_{ij})\right]
     \vec{\sigma}_i \cdot \vec{\sigma}_j 
\nonumber \\
&&   \qquad\qquad +\left[H (m_{\pi} \,r_{ij})
     -\dfrac{\Lambda^3}{m_{\pi}^3} H(\Lambda
\,r_{ij}) \right] S_{ij}
     \Biggr\}  (\vec{\tau}_i \cdot \vec{\tau}_j)
\, .
\end{eqnarray}
$g^2_{\rm ch}/4\pi$ is the chiral coupling constant,
$Y(x)$ is the standard Yukawa function defined by $Y(x)=e^{-x}/x$,
$S_{ij} \, = \, 3 \, ({\vec \sigma}_i \cdot
{\hat r}_{ij}) ({\vec \sigma}_j \cdot  {\hat r}_{ij})
\, - \, {\vec \sigma}_i \cdot {\vec \sigma}_j$ is
the quark tensor operator, and $H(x)=(1+3/x+3/x^2)\,Y(x)$.

Finally, any model imitating QCD should incorporate
confinement. Being a basic term from the spectroscopic point of view
it is negligible for the hadron-hadron interaction. Lattice calculations 
suggest a screening effect on the potential when increasing the interquark 
distance~\cite{Bal01},
\begin{equation}
V_{\rm CON}(\vec{r}_{ij})=\{-a_{c}\,(1-e^{-\mu_c\,r_{ij}})\}(\vec{%
\lambda^c}_{i}\cdot \vec{ \lambda^c}_{j})\, .
\end{equation}
Once perturbative (one-gluon exchange) and nonperturbative (confinement and
chiral symmetry breaking) aspects of QCD have been considered, one ends up with
a quark-quark interaction of the form 
\begin{equation} 
V_{q_iq_j}(\vec{r}_{ij})=\left\{ \begin{array}{ll} 
\left[ q_iq_j=nn \right] \Rightarrow V_{\rm CON}(\vec{r}_{ij})+V_{\rm OGE}(\vec{r}_{ij})+V_{\chi}(\vec{r}_{ij}) &  \\ 
\left[ q_iq_j=cn/cc \right]  \Rightarrow V_{\rm CON}(\vec{r}_{ij})+V_{\rm OGE}(\vec{r}_{ij}) &

\end{array} \right.\,,
\label{pot}
\end{equation}
\begin{table}[b]
\caption{Quark-model parameters.}
\label{tabn}
\begin{tabular}{cc|cc}
\hline
\hline
 $m_{u,d} ({\rm MeV})$    & 313    & $g_{\rm ch}^2/(4\pi)$      & 0.54  \\ 
 $m_c ({\rm MeV})$        & 1752   & $m_\sigma ({\rm fm^{-1}})$ & 3.42 \\ 
 $b ({\rm fm})$           & 0.518  & $m_\pi ({\rm fm^{-1}})$    & 0.70  \\ 
 $b_c ({\rm fm})$         & 0.6    & $\Lambda ({\rm fm^{-1}})$  & 4.2  \\ 
 $\hat r_0$ (MeV fm)      & 28.170 & $a_c$ (MeV)                & 230 \\ 
 $\mu_c$ (fm$^{-1}$)      & 0.70   &                            &  \\ 
\hline
\hline
\end{tabular}
\end{table}
where $n$ stands for the light quarks $u$ and $d$.
Notice that for the particular case of heavy quarks ($c$ or $b$) chiral symmetry is
explicitly broken and therefore boson exchanges do not contribute.
The parameters of the model are the same that have been used for the
study of the $N\bar D$ system~\cite{Cav12} and for completeness are quoted in Table~\ref{tabn}.
The model guarantees a nice description of the light~\cite{Vag05} and charmed~\cite{Val08} baryon spectra.

In order to derive the $B_n B_m\to B_k B_l$ interaction from the
basic $qq$ interaction defined above, we use a Born-Oppenheimer
approximation. Explicitly, the potential is calculated as follows,
\begin{equation}
V_{B_n B_m (L \, S \, T) \rightarrow B_k B_l (L^{\prime}\, S^{\prime}\, T)} (R) =
\xi_{L \,S \, T}^{L^{\prime}\, S^{\prime}\, T} (R) \, - \, \xi_{L \,S \,
T}^{L^{\prime}\, S^{\prime}\, T} (\infty) \, ,  \label{Poten1}
\end{equation}
\noindent where
\begin{equation}
\xi_{L \, S \, T}^{L^{\prime}\, S^{\prime}\, T} (R) \, = \, {\frac{{\left
\langle \Psi_{B_k B_l}^{L^{\prime}\, S^{\prime}\, T} ({\vec R}) \mid
\sum_{i<j=1}^{6} V_{q_iq_j}({\vec r}_{ij}) \mid \Psi_{B_n B_m}^{L \, S \, T} ({\vec R%
}) \right \rangle} }{{\sqrt{\left \langle \Psi_{B_k B_l }^{L^{\prime}\,
S^{\prime}\, T} ({\vec R}) \mid \Psi_{B_k B_l }^{L^{\prime}\, S^{\prime}\, T} ({%
\vec R}) \right \rangle} \sqrt{\left \langle \Psi_{B_n B_m }^{L \, S \, T} ({\vec %
R}) \mid \Psi_{B_n B_m }^{L \, S \, T} ({\vec R}) \right \rangle}}}} \, .
\label{Poten2}
\end{equation}
In the last expression the quark coordinates are integrated out keeping $R$
fixed, the resulting interaction being a function of the two-baryon relative 
distance. The wave function $\Psi_{B_n B_m}^{L \, S \, T}({\vec R})$ for the two-baryon
system has been discussed in Sec.~\ref{secIIA}.

\subsection{Integral equations for the two-body systems}
\label{secIIC}
\begin{table}[b]
\caption{$S$ and $D$ wave two-baryon channels contributing to the different isospin-spin $(T,J$) states. See text for details.}
\label{tab2}
\begin{tabular}{c|cccc}
\hline \hline
      & $T=0$                                    &&  $T=1$                                    & $T=2$          \\
\hline
$J=0$ & $\Lambda_c\Lambda_c \, / \, N \Xi_{cc} \, / \, \Sigma_c\Sigma_c$  && $N \Xi_{cc} \, / \, \Lambda_c \Sigma_c$              & $\Sigma_c\Sigma_c$  \\
$J=1$ & $N\Xi_{cc}$                                   && $N \Xi_{cc} \, / \, \Lambda_c\Sigma_c \, / \, \Sigma_c\Sigma_c$    & $-$ \\
\hline\hline
\end{tabular}
\end{table}

To study the possible existence of two-baryon molecular states with two units of charm: $\Lambda_c\Lambda_c$, 
$N\Xi_{cc}$, $\Lambda_c\Sigma_c$, or $\Sigma_c\Sigma_c$, we have solved the Lippmann-Schwinger equation 
for negative energies looking at the Fredholm determinant $D_F(E)$ at zero 
energy~\cite{Gar87}. If there are no interactions then $D_F(0)=1$, 
if the system is attractive then $D_F(0)<1$, and if a
bound state exists then $D_F(0)<0$. This method permits
us to obtain robust predictions even for zero-energy bound states, and gives
information about attractive channels that may lodge a resonance in similar systems~\cite{Car09}.
We consider a two-baryon system $B_i B_j$ in a relative $S$ state interacting through a potential $V$ that contains a
tensor force. Then, in general, there is a coupling to the 
$B_i B_j$ $D$ wave. Moreover, the two-baryon system can couple to other 
two-baryon states. We show in Table~\ref{tab2} the  two-baryon
coupled channels in the isospin-spin $(T,J)$ basis.

Thus, if we denote the different two-baryon systems as channel $A_i$,
the Lippmann-Schwinger equation for the baryon-baryon scattering becomes
\begin{eqnarray}
t_{\alpha\beta;TJ}^{\ell_\alpha s_\alpha, \ell_\beta s_\beta}(p_\alpha,p_\beta;E)& = & 
V_{\alpha\beta;TJ}^{\ell_\alpha s_\alpha, \ell_\beta s_\beta}(p_\alpha,p_\beta)+
\sum_{\gamma=A_1,A_2,\cdots}\sum_{\ell_\gamma=0,2} 
\int_0^\infty p_\gamma^2 dp_\gamma V_{\alpha\gamma;TJ}^{\ell_\alpha s_\alpha, \ell_\gamma s_\gamma}
(p_\alpha,p_\gamma) \nonumber \\
& \times& \, G_\gamma(E;p_\gamma)
t_{\gamma\beta;TJ}^{\ell_\gamma s_\gamma, \ell_\beta s_\beta}
(p_\gamma,p_\beta;E) \,\,\,\, , \, \alpha,\beta=A_1,A_2,\cdots \,\, ,
\label{eq0}
\end{eqnarray}
where $t$ is the two-body scattering amplitude, $T$, $J$, and $E$ are the
isospin, total angular momentum and energy of the system,
$\ell_{\alpha} s_{\alpha}$, $\ell_{\gamma} s_{\gamma}$, and
$\ell_{\beta} s_{\beta }$
are the initial, intermediate, and final orbital angular momentum and spin, respectively,
 and $p_\gamma$ is the relative momentum of the
two-body system $\gamma$. The propagators $G_\gamma(E;p_\gamma)$ are given by
\begin{equation}
G_\gamma(E;p_\gamma)=\frac{2 \mu_\gamma}{k^2_\gamma-p^2_\gamma + i \epsilon} \, ,
\end{equation}
with
\begin{equation}
E=\frac{k^2_\gamma}{2 \mu_\gamma} \, ,
\end{equation}
where $\mu_\gamma$ is the reduced mass of the two-body system $\gamma$.
For bound-state problems $E < 0$ so that the singularity of the propagator
is never touched and we can forget the $i\epsilon$ in the denominator.
If we make the change of variables
\begin{equation}
p_\gamma = d\frac{1+x_\gamma}{1-x_\gamma},
\label{eq2}
\end{equation}
where $d$ is a scale parameter, and the same for $p_\alpha$ and $p_\beta$, we can
write Eq.~(\ref{eq0}) as
\begin{eqnarray}
t_{\alpha\beta;TJ}^{\ell_\alpha s_\alpha, \ell_\beta s_\beta}(x_\alpha,x_\beta;E)& = & 
V_{\alpha\beta;TJ}^{\ell_\alpha s_\alpha, \ell_\beta s_\beta}(x_\alpha,x_\beta)+
\sum_{\gamma=A_1,A_2,\cdots}\sum_{\ell_\gamma=0,2} 
\int_{-1}^1 d^2\left(\frac{1+x_\gamma}{1-x_\gamma} \right)^2 \,\, \frac{2d}{(1-x_\gamma)^2} \,
dx_\gamma \nonumber \\
&\times & V_{\alpha\gamma;TJ}^{\ell_\alpha s_\alpha, \ell_\gamma s_\gamma}
(x_\alpha,x_\gamma) \, G_\gamma(E;p_\gamma) \,
t_{\gamma\beta;TJ}^{\ell_\gamma s_\gamma, \ell_\beta s_\beta}
(x_\gamma,x_\beta;E) \, .
\label{eq3}
\end{eqnarray}
We solve this equation by replacing the integral from $-1$ to $1$ by a
Gauss-Legendre quadrature which results in the set of
linear equations
\begin{equation}
\sum_{\gamma=A_1,A_2,\cdots}\sum_{\ell_\gamma=0,2}\sum_{m=1}^N
M_{\alpha\gamma;TJ}^{n \ell_\alpha s_\alpha, m \ell_\gamma s_\gamma}(E) \, 
t_{\gamma\beta;TJ}^{\ell_\gamma s_\gamma, \ell_\beta s_\beta}(x_m,x_k;E) =  
V_{\alpha\beta;TJ}^{\ell_\alpha s_\alpha, \ell_\beta s_\beta}(x_n,x_k) \, ,
\label{eq4}
\end{equation}
with
\begin{eqnarray}
M_{\alpha\gamma;TJ}^{n \ell_\alpha s_\alpha, m \ell_\gamma s_\gamma}(E)
& = & \delta_{nm}\delta_{\ell_\alpha \ell_\gamma} \delta_{s_\alpha s_\gamma}
- w_m d^2\left(\frac{1+x_m}{1-x_m}\right)^2 \frac{2d}{(1-x_m)^2} \nonumber \\
& \times & V_{\alpha\gamma;TJ}^{\ell_\alpha s_\alpha, \ell_\gamma s_\gamma}(x_n,x_m) 
\, G_\gamma(E;{p_\gamma}_m),
\label{eq5}
\end{eqnarray}
and where $w_m$ and $x_m$ are the weights and abscissas of the Gauss-Legendre
quadrature while ${p_\gamma}_m$ is obtained by putting
$x_\gamma=x_m$ in Eq.~(\ref{eq2}).
If a bound state exists at an energy $E_B$, the determinant of the matrix
$M_{\alpha\gamma;TJ}^{n \ell_\alpha s_\alpha, m \ell_\gamma s_\gamma}(E_B)$ 
vanishes, i.e., $\left|M_{\alpha\gamma;TJ}(E_B)\right|=0$.

\section{Results and discussion}
\label{secIII}
\begin{figure}[t]
\includegraphics[width=7cm]{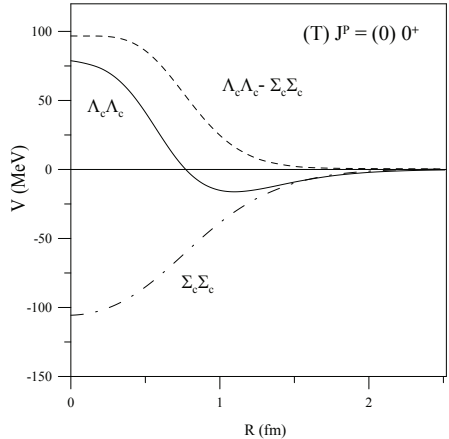}
\caption{Different two-body potentials contributing to the $(T)J^P=(0)0^+$
channel.}
\label{fig3}
\end{figure}
\begin{figure*}[b]
\includegraphics[width=7cm]{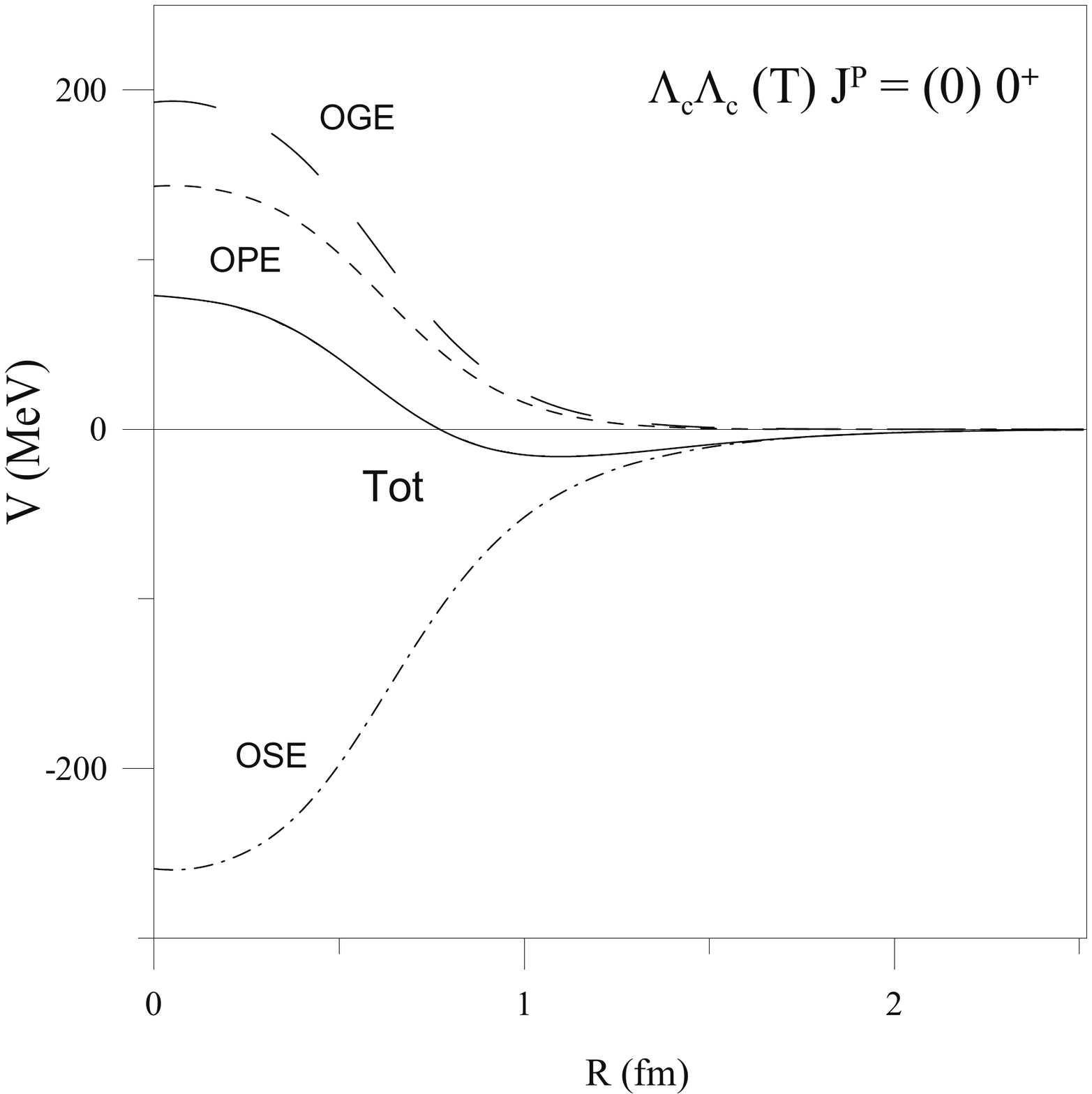}\hspace{+2cm}
\includegraphics[width=7cm]{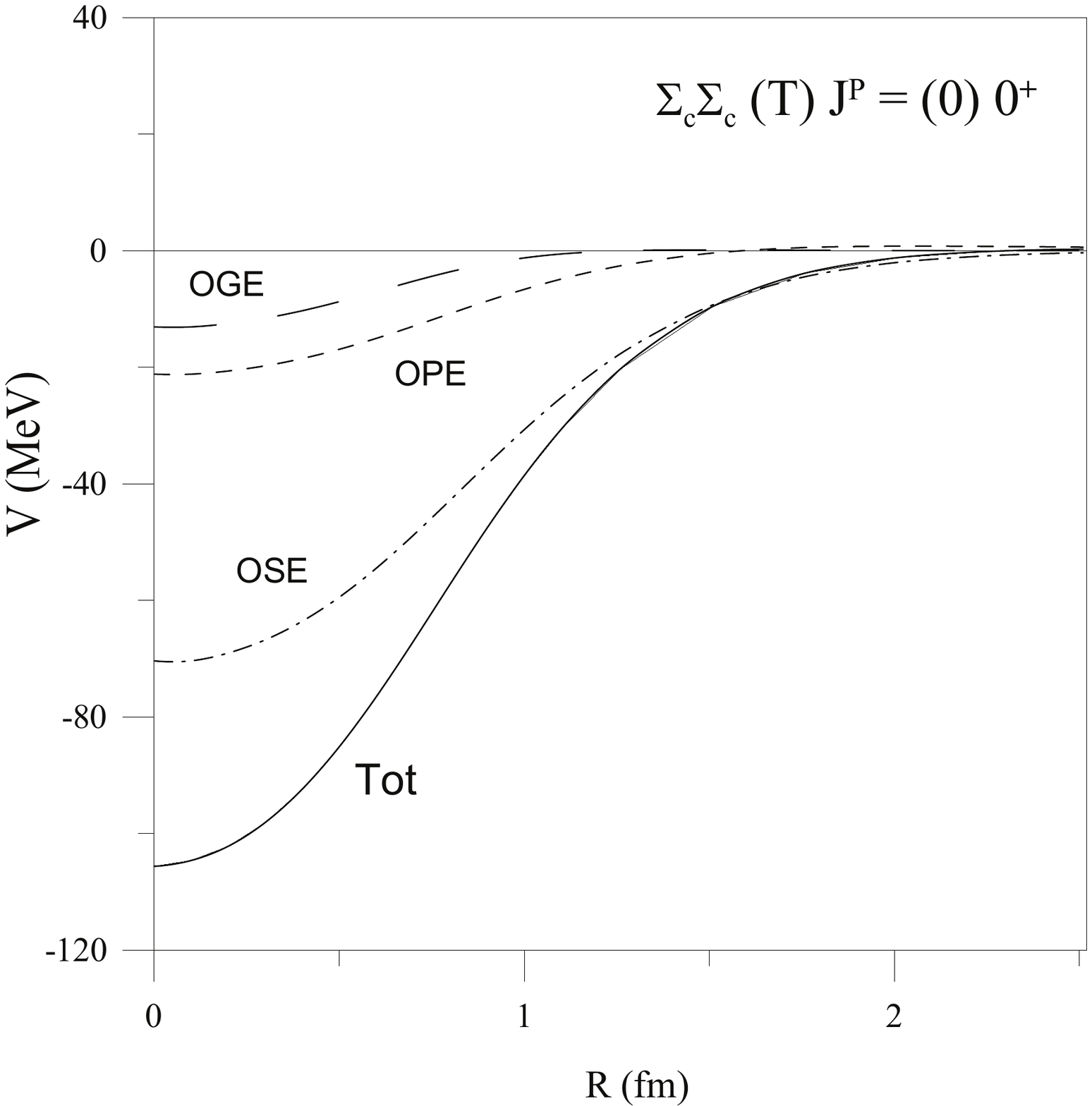}
\caption{Left panel: Contribution of the different terms of the interaction to the
$(T)J^P=(0)0^+$ $\Lambda_c\Lambda_c$ potential. 'OGE' stands for the one-gluon exchange, 
'OPE' for the one-pion exchange, 'OSE' denotes the one-sigma exchange and 'Tot'
represents the total potential. Right panel: Same as the left panel
for the $(T)J^P=(0)0^+$ $\Sigma_c\Sigma_c$ potential. }
\label{fig4}
\end{figure*}
We will first discuss the interactions derived with
the CCQM, centering our attention in the most interesting channel, the
flavor singlet channel $(T)J^P=(0)0^+$ with the quantum numbers of the $\Lambda_c\Lambda_c$
state. This channel might lodge a charmed $H$-like dibaryon. We show in Fig.~\ref{fig3} the diagonal and
transition central potentials contributing to the $(T)J^P=(0)0^+$ state.
It is important to note that the $\Lambda_c\Lambda_c$ system is decoupled from the
closest two-baryon threshold, the $N\Xi_{cc}$ state, that in the case of the
strange $H$ dibaryon becomes relevant for its possible bound or resonant character~\cite{Ino12}. 
The binding of the $(T)J^P=(0)0^+$ state would then require a stronger attraction
in the diagonal channels or a stronger coupling to the heavier $\Sigma_c\Sigma_c$ state,
that as we will discuss below is not fulfilled.

In Fig.~\ref{fig4} we have separated the contributions of the different terms
in Eq.~(\ref{pot}) to the two diagonal interactions. As can be seen,
the $\Lambda_c\Lambda_c$ potential is the most repulsive one. It becomes repulsive
at short-range partially due to the reduction of the normalization kernel (see Fig.~\ref{fig2}).
The OGE and OPE can only give contributions through quark-exchange diagrams due to the
color-spin-isospin structure of the antisymmetry operator~\cite{Shi84,Val05}. They 
generate short-range repulsion that it is 
compensated at intermediate distances by the attraction coming from the scalar exchange, with a longer range. 
Thus, the total potential becomes slightly attractive at intermediate distances but 
repulsive at short range.
In the $\Sigma_c\Sigma_c$ interaction, the presence of a direct (without 
simultaneous quark-exchange) contribution of the OPE and the opposite sign of most part of the exchange
diagrams, generates an overall attractive potential. This is rather similar
to the situation in the strange sector but with the absence of the one-kaon
exchange potential, what gives rise to a less attractive interaction. Regarding the
character of the interaction, similar results were obtained in Ref.~\cite{Hua14}
within the quark delocalization color screening model (QDCSM). It is important to
note at this point the difference with hadronic potential models as those of 
Refs.~\cite{Lee11,Meg11}. As can be seen in Fig. 3(a) of Ref.~\cite{Lee11},
the $(T)J^P=(0)0^+$ $\Lambda_c\Lambda_c$ potential is attractive due to the absence of quark-exchange
contributions and the dominance of the attraction of the scalar exchange potential. In spite
of being attractive, the central potential alone is not enough to generate a bound state. In 
Ref.~\cite{Meg11} they only consider the hadronic one-pion exchange and
then, the $\Lambda_c\Lambda_c$ interaction is zero. Thus all possible attraction
comes generated by the coupling to larger mass channels.

As mentioned above, when comparing with the similar problem
in the strange sector an important difference arises,
the absence of the $\Lambda_c\Lambda_c \leftrightarrow N\Xi_{cc}$ coupling. As a 
consequence the mass difference between the two coupled channels in 
the $(T)J^P=(0)0^+$ partial wave, $\Lambda_c\Lambda_c$ and $\Sigma_c\Sigma_c$, is 
much larger than in the strange sector, making the 
coupled channel effect less important.
Let us note that in the strange sector $M(N\Xi) - M(\Lambda\Lambda)=$ 25 MeV and 
$M(\Sigma\Sigma) - M(\Lambda\Lambda)=$ 154 MeV, this is why the $N\Xi$ channel
plays a relevant role for the $\Lambda\Lambda$ system~\cite{Ino12}, as well as why the $N\Sigma$
state is relevant for the $N\Lambda$ system~\cite{Gar07}. In the charmed sector
the closest channel coupled to $\Lambda_c\Lambda_c$ in the $(T)J^P=(0)0^+$ state is
$\Sigma_c\Sigma_c$, 338 MeV above. This energy difference is similar to the 
$N\Delta- NN$ mass difference, the coupled channel effect being still important
although it may not proceed through the central terms due to angular momentum
selection rules~\cite{Val95}. Heavier channels play a minor role, 
as it occurs with the $\Delta\Delta$ channel, 584 MeV above the $NN$ 
threshold~\cite{Val95}. In the present case the coupling to the closest
channel $\Sigma_c\Sigma_c$ proceeds through the central potential and one does not 
expect higher channels, as $\Sigma_c^*\Sigma_c^*$ 468 MeV
above the $\Lambda_c\Lambda_c$ threshold, to play a relevant role in quark-model
descriptions as shown in the QDCSM model of Ref.~\cite{Hua14}. 
The situation seems to be a bit different in hadronic models where the non-central
potentials are not regularized by the quark-model wave function~\cite{Meg11,Liz12}.
Let us finally note that the coupling between the $\Lambda_c\Lambda_c$ 
and $\Sigma_c\Sigma_c$ channels comes mainly given by quark-exchange
effects and the direct one-pion exchange potential. Thus, it becomes a little
bit stronger than in hadronic theories based on the one-pion exchange potential~\cite{Meg11}.
The resulting interaction is rather similar to that in the strange sector, as can be seen by comparing with
Fig. 1(b) of Ref.~\cite{Car12}, the main difference coming from the behavior of the normalization kernel
at short distances.
\begin{figure*}[t]
\includegraphics[width=7cm]{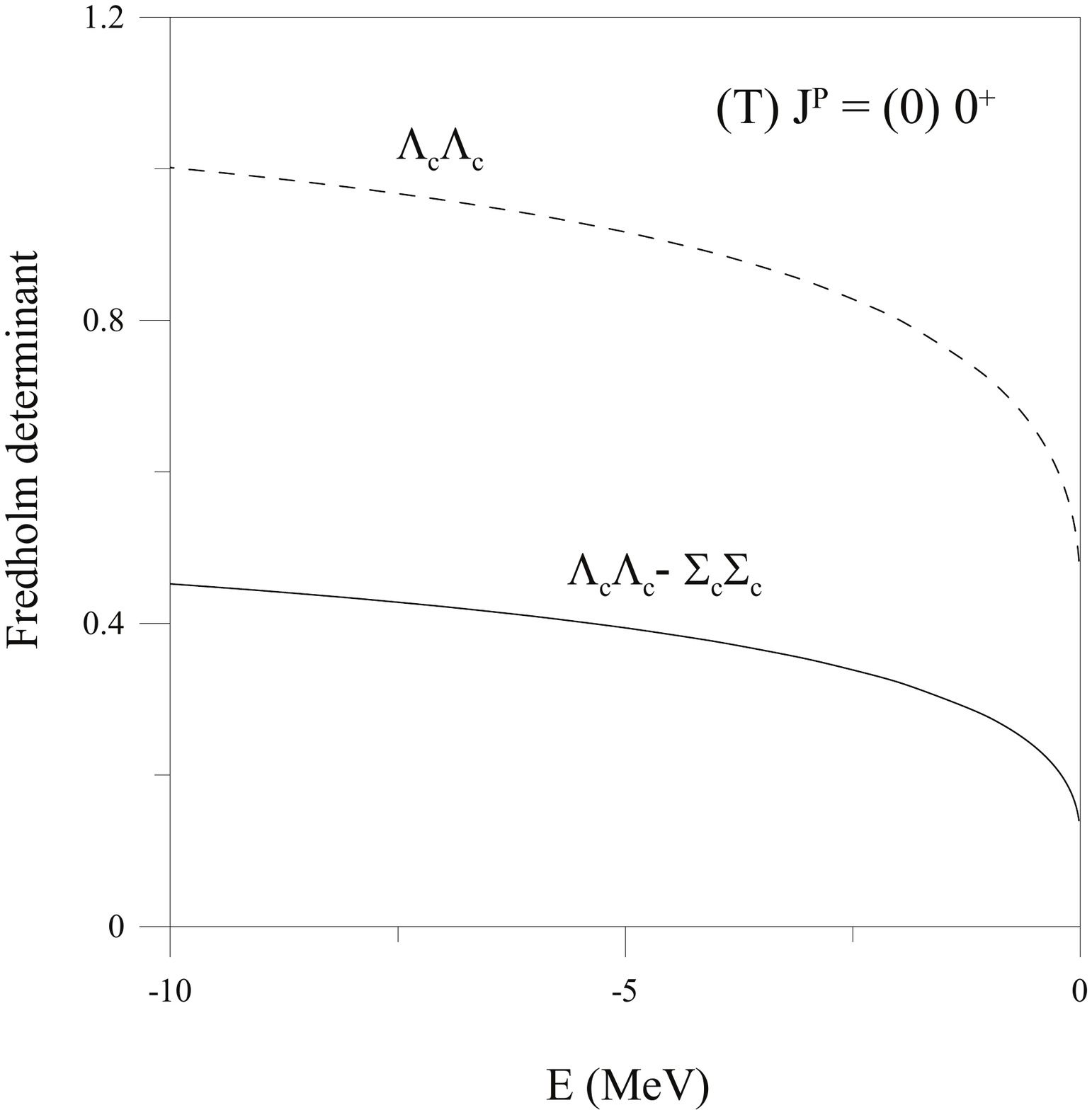}\hspace{+2cm}
\includegraphics[width=7cm]{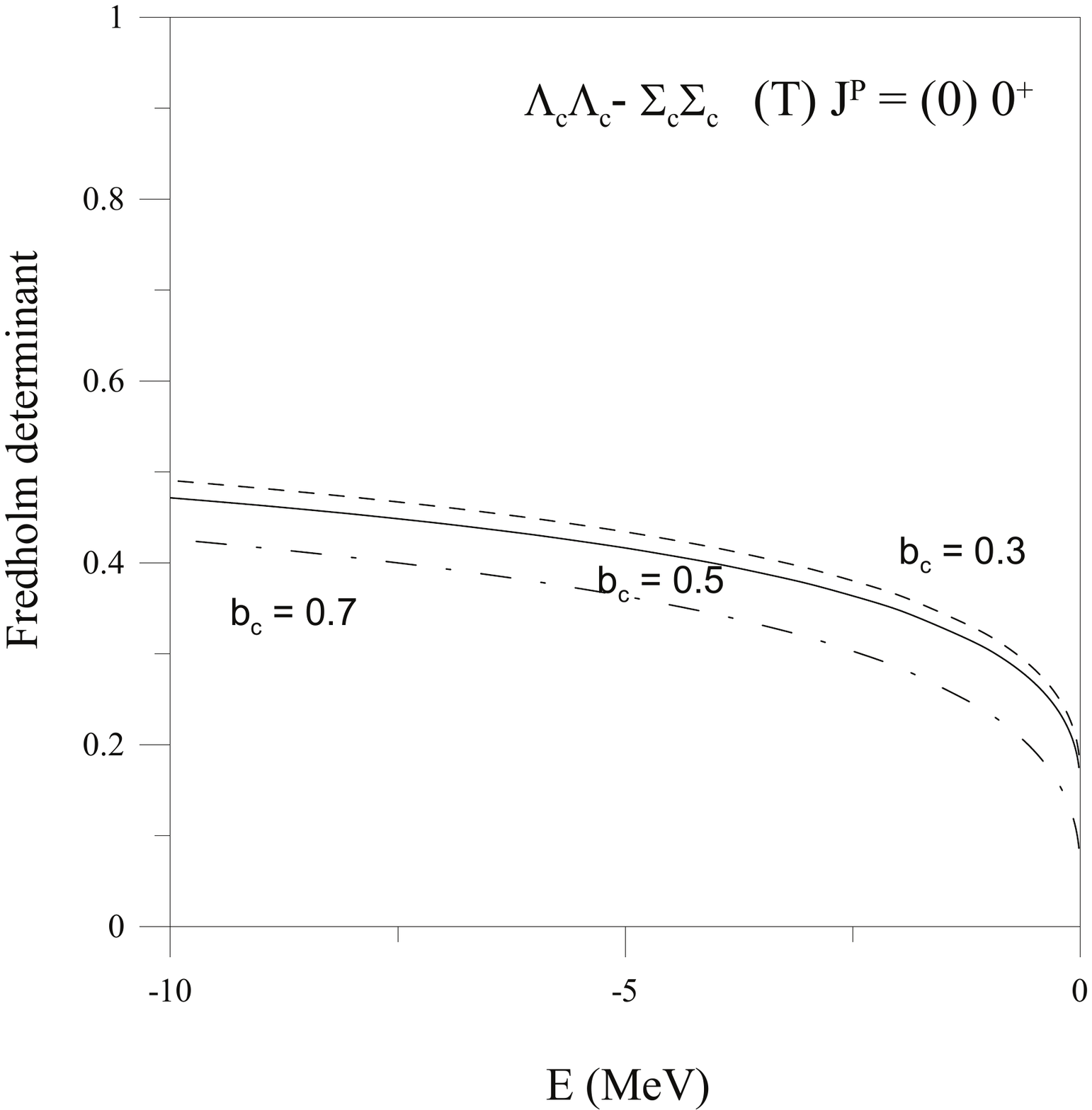}
\caption{Left panel: Fredholm determinant of the $(T)J^P=(0)0^+$ channel. The dashed line
only considers the $\Lambda_c\Lambda_c$ state, whereas the solid line includes the
coupling to the $\Sigma_c\Sigma_c$ state.
Right panel: Fredholm determinant of the $(T)J^P=(0)0^+$ coupled channel state for different
values of $b_c$ in fm. See text for details.}
\label{fig5}
\end{figure*}
\begin{figure*}[b]
\includegraphics[width=7cm]{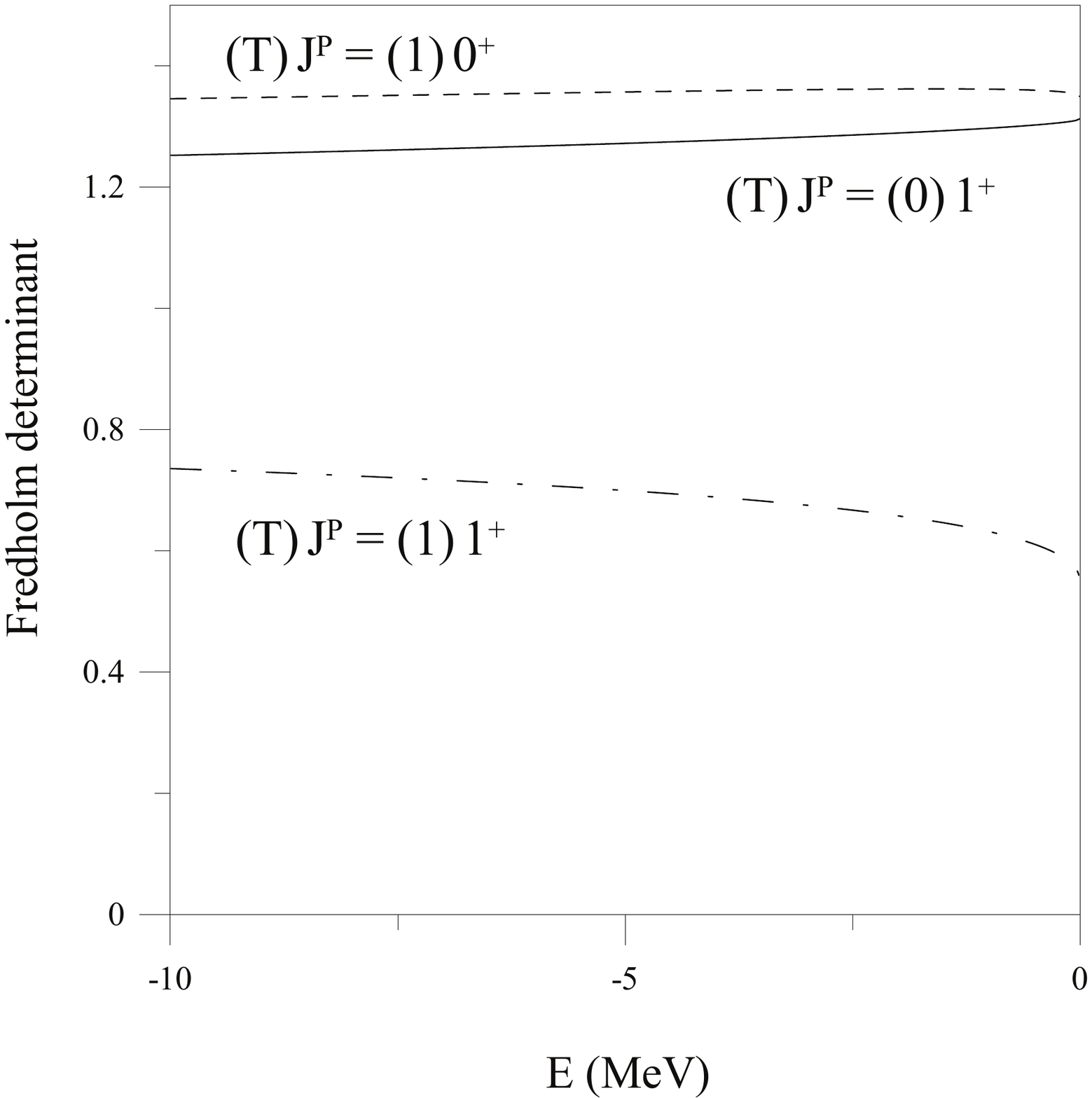}\hspace{+2cm}
\includegraphics[width=7cm]{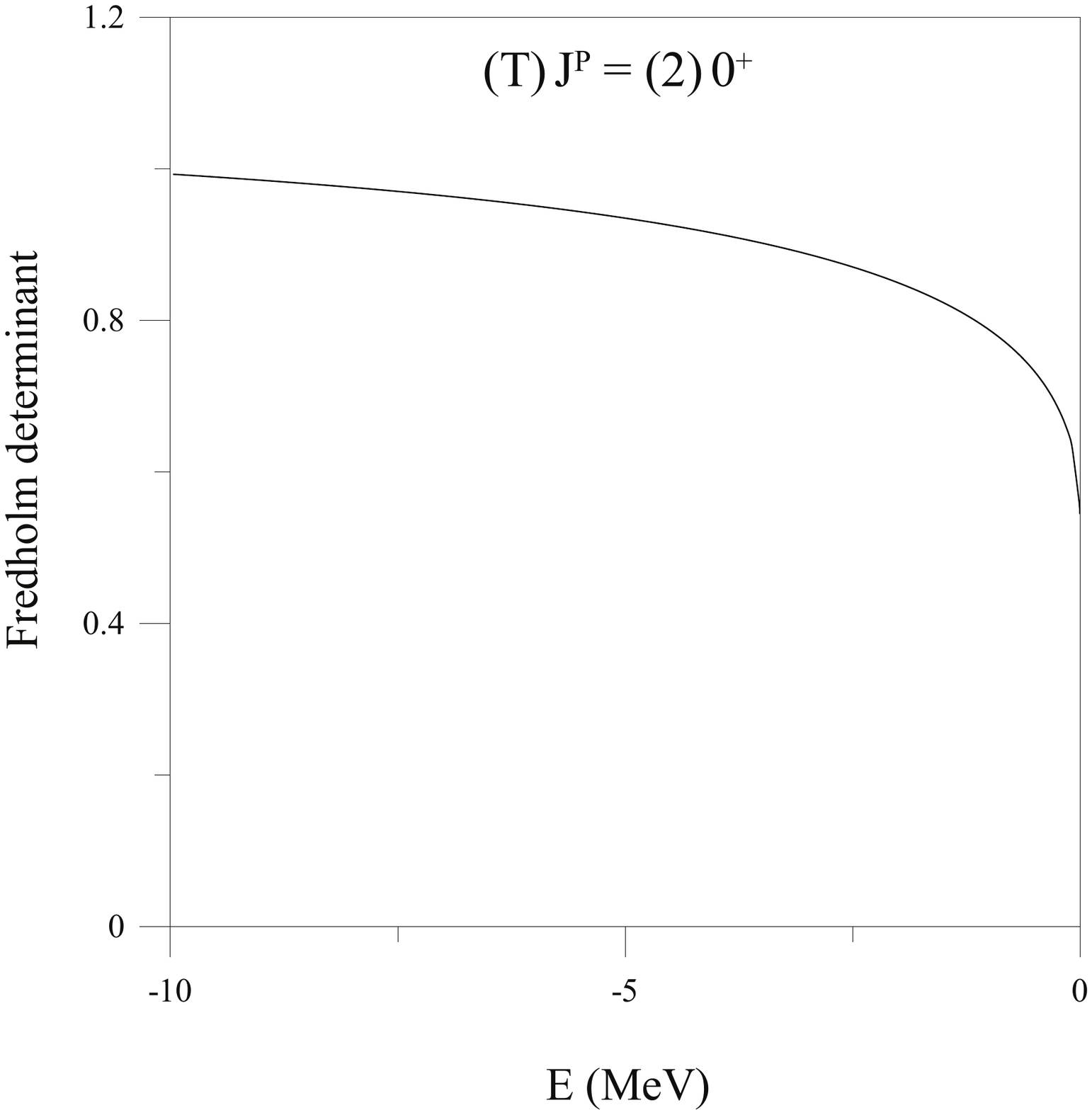}
\caption{Left panel: Fredholm determinant of the $(T)J^P$ channels where the $N\Xi_{cc}$
state is the lowest threshold.
Right panel: Fredholm determinant of the $(T)J^P$ channel where the $\Sigma_c\Sigma_c$
state is the lowest threshold.}
\label{fig6}
\end{figure*}

With these ideas in mind and following
the procedure described in Sec.~\ref{secIIC} we have performed a full coupled channel
calculation of the $(T)J^P=(0)0^+$ state. The results are shown in Fig.~\ref{fig5}.
In the left panel we show by the dashed line the Fredholm determinant of the
$\Lambda_c\Lambda_c$ channel alone. The Fredholm determinant is large, indicating 
a barely small attractive interaction. When the coupling
to the heavier $\Sigma_c\Sigma_c$ channel is included, solid line in Fig.~\ref{fig5}, 
the system gains attraction, but it is not enough as to get a bound state. 
The main uncertainty when determining the baryon-baryon interaction in quark 
models with charmed baryons would be the harmonic oscillator parameter of the charm 
quark. We have explored the results for different values of $b_c$. 
The results are shown in the right panel of Fig.~\ref{fig5} and as can be seen in no 
case the $(T)J^P=(0)0^+$ state would become bound. Note that in Ref.~\cite{Car11} it was argued
that the smaller values of $b_c$ are preferred to get consistency with 
calculations based on infinite expansions, as hyperspherical harmonic 
expansions~\cite{Vij09}, where the quark wave function is not postulated. 
This also agrees with simple harmonic oscillator relations $b_c=b_n\sqrt{\frac{m_n}{m_c}}$.
The smaller values of $b_c$ give rise to the less attractive results.
For the larger values of $b_c$, if a loosely bound state could be generated,
the electromagnetic repulsion arising in the $(T)J^P=(0)0^+$ channel
due to the electric charge of the $\Lambda^+_c$ might dismantle the bound state.

Thus, without the strong transition potentials reported in the QDCSM model of Ref.~\cite{Hua14} or 
the strong tensor couplings occurring in the hadronic one-pion exchange models of Refs.~\cite{Meg11,Liz12}, 
it seems difficult to get a bound state in this system. We have 
recently illustrated within the quark model~\cite{Vij14} how the coupled 
channel effect between channels with an almost negligible interaction in the 
lower mass channel works for generating bound states. Although for the four-quark problem,
in this reference it is demonstrated (see Fig. 2 of Ref.~\cite{Vij14}) how when the
thresholds mass difference increases, the effect of the coupled channel diminishes,
which is an unavoidable consequence of having the same hamiltonian to describe the 
hadron masses and the hadron-hadron interactions.

We have also analyzed the other $(T)J^P$ channels shown in Table~\ref{tab2} with similar conclusions, 
the weak interaction in the charm sector and the absence of channel
coupling between close mass channels works against the possibility of having 
dibaryons with two units of charm. For the sake of completeness we have calculated the
Fredholm determinant for all cases and it is shown in Fig.~\ref{fig6}.

One should finally note that the problem of double 
heavy dibaryons has also been approached in the literature by means of 
six-quark calculations. The group of Grenoble~\cite{Lea95} addressed this 
problem within a pure chromomagnetic interaction obtaining several candidates 
to be bound. There also interesting results based on relativistic 
six-quark equations constructed in the framework of the dispersion relation 
technique~\cite{Ger12} with a rich spectroscopy of double charmed and beauty heavy
dibaryons. Future experimental results will help to scrutinize among 
the different models, and in this way to improve our phenomenological
understanding of QCD in the highly non-perturbative low-energy regime.
This challenge could only be achieved by means of a cooperative experimental
and theoretical effort.

\section{Summary}
\label{secIV}
In short summary, we have studied the baryon-baryon interaction with two units of charm
making use of a chiral constituent quark model tuned in the description of the baryon 
and meson spectra as well as the $NN$ interaction. Several effects conspire against the
existence of a loosely bound state resembling the 
$H$ dibaryon. First, the interaction is weaker than in
the strange sector. Second, there is no coupling between close mass thresholds,
like $\Lambda_c\Lambda_c \leftrightarrow N\Xi_{cc}$, the closest threshold being more than 300 MeV
above. Finally, the existence of a weak attraction may be killed by the
electromagnetic repulsion absent in the strange sector.
Thus, our results point to the nonexistence of low-energy dibaryons with two
units of charm and in particular, to the nonexistence of a stable charmed $H$-like dibaryon.
Given that the interaction in the $(T)J^P=(0)0^+$ is attractive, this state may
appear as a resonance above the $\Lambda_c\Lambda_c$ threshold.

Weakly bound states are usually very sensitive to potential details and therefore
theoretical investigations with different phenomenological models are highly desirable.
The existence of these states could be scrutinized in the future at the LHC, J-PARC and
RHIC providing a great opportunity for extending our knowledge to some unreached 
part in our matter world.

\section{Acknowledgments}
This work has been partially funded by the Spanish Ministerio de
Educaci\'on y Ciencia and EU FEDER under Contract No. FPA2013-47443-C2-2-P,
and by the Spanish Consolider-Ingenio 2010 Program CPAN (CSD2007-00042).


\begin{thebibliography}{99}

\bibitem{Esp14} A.~Esposito, A.~L.~Guerrieri, F.~Piccinini, A.~Pilloni, and A.~D.~Polosa,
								Int. J. Mod. Phys. A {\bf 30}, 1530002 (2014). 

\bibitem{Cho08} S.~-K.~Choi {\it et al.} (Belle Collaboration),
								Phys. Rev. Lett. {\bf 100}, 142001 (2008).

\bibitem{Bon12} A.~Bondar {\it et al.} (Belle Collaboration),
								Phys. Rev. Lett. {\bf 108}, 122001 (2012).

\bibitem{Jaf77} R.~L.~Jaffe, 
								Phys. Rev. Lett. {\bf 38}, 195 (1977); {\bf 38}, 617(E) (1977).

\bibitem{Lee11} N.~Lee, Z.~-G.~Luo, X.~-L.~Chen, and S.~-L.~Zhu,
								Phys. Rev. D {\bf 84}, 014031 (2011).
								
\bibitem{Meg11} W.~Meguro, Y.~-R.~Liu, and M.~Oka,
								Phys. Lett. B {\bf 704}, 547 (2011).
								
\bibitem{Liz12} N.~Li and S.~-L.~Zhu,
								Phys. Rev. D {\bf 86}, 014020 (2012).
								
\bibitem{Oka13} M.~Oka,
								Nucl. Phys. A {\bf 914}, 447 (2013).
							
\bibitem{Hua14}	H.~Huang, J.~Ping, and F.~Wang,
								Phys. Rev. C {\bf 89}, 035201 (2014).

\bibitem{Ric15} J.~-M.~Richard, Q.~Wang, and Q.~Zhao,
								Phys. Rev. C {\bf 91}, 014003 (2015).

\bibitem{Val05} A.~Valcarce, H.~Garcilazo, F.~Fern\'andez, and P.~Gonz\'alez,
								Rep. Prog. Phys. {\bf 68}, 965 (2005).

\bibitem{Vij05} J.~Vijande, F.~Fern\'andez, and A.~Valcarce,
								J. Phys. G {\bf 31}, 481 (2005). 

\bibitem{Vag05} A.~Valcarce, H.~Garcilazo, and J.~Vijande,
								Phys. Rev. C {\bf 72}, 025206 (2005).
								
\bibitem{Val08} A.~Valcarce, H.~Garcilazo, and J.~Vijande,					
								Eur. Phys. J. A {\bf 37}, 217 (2008).

\bibitem{Car09} T.~Fern\'andez-Caram\'es, A.~Valcarce, and J.~Vijande,
								Phys. Rev. Lett. {\bf 103}, 222001 (2009).

\bibitem{Pan09} U.~Wiedner ({$\bar {\rm P}$}ANDA Collaboration), 
								Prog. Part. Nucl. Phys. {\bf 66}, 477 (2011).

\bibitem{Dov77} C.~B.~Dover and S.~H.~Kahana, 
								Phys. Rev. Lett. {\bf 39}, 1506 (1977).

\bibitem{Val97} A.~Valcarce, F.~Fern\'andez, and P.~Gonz\'alez,
								Phys. Rev. C {\bf 56}, 3026 (1997).

\bibitem{Ruj75} A.~de~R\'ujula, H.~Georgi, and S.~L.~Glashow,
								Phys. Rev. D {\bf 12}, 147 (1975).

\bibitem{Bal01} G.~S.~Bali, 
								Phys. Rep. {\bf 343}, 1 (2001).
								
\bibitem{Cav12} T.~F.~Caram\'es and A.~Valcarce,
								Phys. Rev. D {\bf 85}, 094017 (2012).

\bibitem{Gar87} H.~Garcilazo, 
								J. Phys. G {\bf 13}, L63 (1987).

\bibitem{Ino12}	T.~Inoue, S.~Aoki, T.~Doi, T.~Hatsuda, Y.~Ikeda, N.~Ishii, 
								K.~Murano, H.~Nemura, and K.~Sasaki (HAL QCD Collaboration),
								Nucl. Phys. A {\bf 881},28 (2012).

\bibitem{Shi84} K.~Shimizu,
								Phys. Lett. B {\bf 148}, 418 (1984).

\bibitem{Gar07} H.~Garcilazo, A.~Valcarce, and T.~Fern\'andez-Caram\'es,
								Phys. Rev. C {\bf 76}, 034001 (2007); {\bf 75}, 034002 (2007).

\bibitem{Val95} A.~Valcarce, A.~Faessler, and F.~Fern\'andez,
								Phys. Lett. B {\bf 345}, 367 (1995).
								
\bibitem{Car12} T.~F.~Caram\'es and A.~Valcarce,
								Phys. Rev. C {\bf 85}, 045202 (2012).
							
\bibitem{Car11} T.~F.~Caram\'es, A.~Valcarce, and J.~Vijande,
								Phys. Lett. B {\bf 699}, 291 (2011).

\bibitem{Vij09} J.~Vijande, A.~Valcarce, and N.~Barnea, 
								Phys. Rev. D {\bf 79}, 074010 (2009).

\bibitem{Vij14} J.~Vijande and A.~Valcarce,
								Phys. Lett. B {\bf 736}, 325 (2014).

\bibitem{Lea95} J.~Leandri and B.~Silvestre-Brac,
								Phys. Rev. D {\bf 51}, 3628 (1995).

\bibitem{Ger12} S.~M.~Gerasyuta and E.~E.~Matskevich,
								Int. J. Mod. Phys. E {\bf 21}, 1250058 (2012).
								
\end{thebibliography}
\end{document}